\PassOptionsToPackage{svgnames}{xcolor}
\documentclass[journal]{vgtc}

\onlineid{8416}

\vgtccategory{Research}

\vgtcpapertype{Theoretical \& Empirical}

\usepackage{multirow}
\usepackage{algorithm}
\usepackage{algpseudocode}
\usepackage{csquotes}

\usepackage{tikz}
\usepackage{graphicx}
\usetikzlibrary{positioning, arrows.meta}

\title{Detecting and Understanding Ambiguity in Visualizations with Role-Playing Vision Language Model Interpretations}

\author{%
  Garrick Chan,
  \authororcid{Yun-Hsin Kuo}{0009-0000-1891-8993},
  \authororcid{Sam Yu-Te Lee}{0009-0000-2629-3954},
  Yinuo Tang, and
  Kwan-Liu Ma
}   

\authorfooter{
  \item
  	All authors are with University of California, Davis.
  	E-mail: \{garchan,yskuo,ytlee,fptang,klma\}@ucdavis.edu
}

\abstract{%
    Communicative visualizations can support multiple chart-grounded takeaways, including divergent or conflicting ones; we refer to this variation as interpretive ambiguity. Such ambiguity can undermine a visualization's communicative goal when the supported takeaways diverge from its intended message. Evaluating visualizations with target audiences can surface this variation, but such evaluations can be costly and difficult to scale. In this work, we explore persona-conditioned Vision Language Models (VLMs) as a complementary evaluation tool for interpretive ambiguity. For a given visualization, our pipeline uses topic-relevant personas to elicit candidate contextual takeaways supported by interpretations of perceived visual patterns, such as trends and clusters. We derive a Target--Pattern--Scope framework for comparing these interpretations, use LLM judges to assign Compatible, Disjoint, or Opposing ambiguity labels to statement pairs, and summarize their distribution in an ambiguity profile. To improve chart grounding, we benchmark candidate VLMs on chart comprehension and select the highest-performing model for each chart type. We evaluate statement-pair labeling against human-authored annotations and apply the approach to real-world, public-facing visualizations curated from established datasets. Through case studies, we examine how perceived visual patterns relate to contextual takeaways and how a visualization redesign changes the sampled interpretations. Our findings show how persona-conditioned VLMs can complement human feedback by generating candidate contextual takeaways and surfacing potential ambiguity for further inspection.

}


\renewcommand{\manuscriptnotetxt}{}

\graphicspath{{figures/}{pictures/}{images/}{./}} 

\usepackage{tabu}                      
\usepackage{booktabs}                  
\usepackage{lipsum}                    
\usepackage{mwe}                       
\usepackage{svg}

\usepackage{mathptmx}                  

\begin{document}


\definecolor{AgreedColor}{HTML}{4CAF50} 
\definecolor{SubsetColor}{HTML}{59A14F} 
\definecolor{NoOverlapColor}{HTML}{BAB0AC} 
\definecolor{ConflictedColor}{HTML}{E15759} 

\definecolor{TargetColor}{HTML}{3783CA} 
\definecolor{PatternColor}{HTML}{CE6702} 
\definecolor{ScopeColor}{HTML}{6A1CBE} 

\newcommand{\hlE}[1]{\colorbox{AgreedColor!25}{#1}}
\newcommand{\hlN}[1]{\colorbox{SubsetColor!40}{#1}}
\newcommand{\hlD}[1]{\colorbox{NoOverlapColor!50}{#1}}
\newcommand{\hlO}[1]{\colorbox{ConflictedColor!40}{#1}}

\newcommand{\tpsTarget}[1]{{\color{TargetColor}#1}}
\newcommand{\tpsPattern}[1]{{\color{PatternColor}#1}}
\newcommand{\tpsScope}[1]{{\color{ScopeColor}#1}}

\firstsection{Introduction} 

\maketitle
Communicative visualizations are designed to convey data-driven messages from authors to audiences \cite{hullman2011visualization,knoll2025gulf}. Their effectiveness depends not only on accurately representing the data, but also on how audiences interpret what is shown. Yet, a visualization can support more than one chart-grounded takeaway \cite{bearfield2024same,fygenson2025afford}. Viewers may attend to different subsets of the data, describe the same pattern at different scopes, or frame the same comparison in opposing ways. We refer to this phenomenon as \emph{interpretive ambiguity}: variation among the chart-grounded takeaways that viewers can derive from the same visualization. When these takeaways diverge from the intended message, ambiguity may undermine the visualization's communicative goal. Identifying such possibilities during design can help authors determine when a visualization needs clarification or further evaluation with its intended audience.

Visualization research has long examined misleading design practices and their effects on interpretation \cite{huff1954,tufte1986,lo2022misinformed}. Interpretive ambiguity raises a related but distinct concern. Misleadingness asks whether a visualization's design or framing steers viewers toward an improper interpretation; interpretive ambiguity asks whether the visualization supports multiple takeaways. A chart may therefore be ambiguous without exhibiting an obvious design violation, and a misleading chart need not produce varied interpretations. Our definition also differs from ambiguity as an unclear mapping between data and visual encodings \cite{lan2024came}. Instead, it builds on evidence that viewers can derive different takeaways from the same chart based on their prior beliefs, perspectives, and interpretive processes \cite{holder2024polarizing,bearfield2024same,hullman2011visualization}.

The most direct way to surface such variation is to evaluate a visualization with members of its intended audience. Interviews, user studies, and crowdsourcing can provide valuable evidence about audience interpretation, but they can be costly to conduct repeatedly and make it difficult to recruit participants spanning a broad range of relevant perspectives \cite{elmqvist2012patterns,borgo2018information}. We therefore investigate persona-conditioned Vision Language Models (VLMs) as a complementary method for eliciting candidate interpretations at scale. The personas serve as diagnostic probes, not as substitutes for human participants or representative samples of an audience. Our goal is to identify potential ambiguity that an author may wish to investigate, not to predict how actual audiences will interpret a visualization in practice.

Our approach first benchmarks candidate VLMs on chart comprehension and selects the highest-performing model for each chart type. Given a visualization, our pipeline constructs a chart profile and generates questions about its visible patterns, retrieves relevant but diverse personas from PersonaHub \cite{personahub}, and elicits a contextual takeaway and supporting interpretations of perceived visual patterns from each persona in response to the same question. It then compares these interpretations pairwise and summarizes their ambiguity-label distribution in an \emph{ambiguity profile}. This profile does not assign a definitive binary judgment that a visualization is or is not ambiguous; instead, it surfaces differences that visualization authors can examine during design.

Through qualitative analysis of the generated interpretations of perceived visual patterns, we identified three recurring components: the \emph{Target} being described, the claimed \emph{Pattern}, and the \emph{Scope} over which that claim applies. We use these Target--Pattern--Scope components to assign each pair an ambiguity label: \hlN{Compatible}, \hlD{Disjoint}, or \hlO{Opposing}. Compatible pairs express mutually consistent readings, Disjoint pairs foreground different aspects of a chart, and Opposing pairs convey conflicting readings. Disjoint and Opposing pairs provide signals of potential ambiguity for an author to investigate further.

This work contributes:
\begin{itemize}[nosep]
    \item The derivation and operationalization of a Target--Pattern--Scope framework for interpretive ambiguity.
    \item A model-agnostic, persona-conditioned VLM pipeline for generating contextual takeaways and ambiguity profiles.
    \item Case studies applying the framework to real-world visualizations.
\end{itemize}
 \label{sec:intro}
\section{Related Work}

Our work draws on three areas: research on viewer-dependent interpretation of visualizations, evaluations of VLM chart comprehension, and persona-conditioned simulation for design evaluation. Together, these areas motivate using VLM personas to elicit and compare plausible interpretations while also defining the limits of such an approach.

\subsection{Interpretive Ambiguity in Visualization}

Visualization research has addressed several distinct forms of ambiguity, including unclear visual mappings and misleading design choices \cite{lo2022misinformed,lan2024came}, as well as underspecified natural-language commands for visualization generation \cite{luo2026nvbench,tian2025chartgpt} and data exploration \cite{gao2015datatone,voigt2022survey,shen2023language}. Our work instead focuses on \emph{interpretive ambiguity}: variation in the chart-grounded takeaways that different viewers can derive from the same visualization.

Interpretive ambiguity can be understood through the lens of cognitive affordances in interaction design \cite{hartson2003cognitive}. Fygenson et al. \cite{fygenson2025afford} propose a three-step framework in which visualization design decisions and viewer characteristics jointly determine what information an individual can extract and the affordance of a particular takeaway. Design decisions include chart type, encodings, annotations, and displayed statistics. Viewer characteristics include prior beliefs \cite{xiong2019curse}, expertise \cite{vazquez2024knowledge}, and physical or cognitive functionality \cite{etzold2026nonvisual,ziemkiewicz2011locus}. Personal and external factors can therefore influence which information viewers notice and how they interpret it \cite{he2024enthusiastic,burns2025intra}. Author framing can also shape interpretation: Lisnic et al. \cite{lisnic2023mislead}, for example, study faithfully plotted charts presented alongside misleading social-media captions.

Our definition builds on the phenomenon observed by Bearfield et al. \cite{bearfield2024same}, in which individuals derive different takeaways from the same chart through different interpretive processes. Such variation is consequential because ambiguous visualizations can reinforce viewers' existing beliefs \cite{li2025confirmation}. Prior work thus establishes both the diversity of viewer-dependent interpretations and the need to bridge the ``gulf of interpretation'' between authors and audiences \cite{knoll2025gulf}. However, systematically eliciting and comparing possible interpretations during visualization design remains difficult at scale, which motivates us to use persona-conditioned VLMs to generate chart-grounded takeaways from varied simulated perspectives and compare the semantic relationships among them.

\subsection{VLM Chart Comprehension}

Chart comprehension requires VLMs to identify visual encodings, extract values, and reason about relationships represented in a visualization. General-purpose VLMs have been applied to tasks such as automated visualization guidance \cite{das2025misvisfix,ahn2025kim,kim2025design} and data extraction\cite{he2026extract}, but evaluations of their chart comprehension show mixed results. Hong et al. \cite{hong2025VisLit} found that VLMs fell short of human performance on the Visualization Literacy Assessment Test (VLAT) \cite{vlat}. On particular chart-question-answering tasks, structured reasoning prompts that first extract chart properties and values can substantially improve accuracy and sometimes exceed human benchmark performance \cite{das2025charts,aziz2026table,zheng2025advancing,lo2024good}.

Other work develops or probes specialized chart-comprehension methods. Dong and Crisan \cite{dong2026attn} use attention visualizations to examine how fine-tuned open-source models answer chart questions, while Liu et al. \cite{liu2025simvecvis} combine chart images with a vector representation during fine-tuning. Moving beyond direct question answering, Blasilli and Angelini \cite{blasilli2026true} assess VLMs' ability to recognize rhetorical techniques and author intent, and Wang et al. \cite{wang2025align} compare takeaways generated by humans and VLMs for bar charts, finding that human takeaways are generally more diverse although VLM outputs are less affected by changes in chart layout. VLM interpretation also remains vulnerable to deliberately deceptive chart designs \cite{mahbub2025perils}.

Our pipeline therefore treats reliable chart comprehension as a prerequisite rather than a solved problem. Beyond reading the chart correctly, our approach must also elicit the different takeaways that viewers with different perspectives might derive from it. To address this second requirement, we draw on prior work in persona-conditioned simulation.


\subsection{Persona-Conditioned Simulation for Design Evaluation}

Language models can produce responses conditioned on assigned roles or persona descriptions~\cite{shanahan2023role,salewski2023impersonation}. Prior work has used this capability to create scalable probes of possible user behavior. Social Simulacra~\cite{park2022socialsimulacra} generates populated prototypes of social computing systems so that designers can explore interactions that may emerge at a larger scale. Such simulations broaden the range of possible user responses that designers can consider during prototyping.

Building on this broader use of simulated user responses, recent systems apply persona conditioning more directly to design evaluation. Benharrak et al.~\cite{benharrak2024personas} use writer-defined AI personas to provide on-demand feedback from different reader perspectives. PosterMate~\cite{shin2025postermate} uses audience-persona agents to provide and reconcile feedback on visual poster designs, while UXCascade~\cite{holter2026uxcascade} aggregates persona-conditioned interaction traces and feedback to help practitioners identify usability issues and consider design changes. These systems show how persona-conditioned models can generate additional candidate perspectives for consideration during early design iterations.

Persona-conditioned outputs, however, should not be treated as representative human responses. Hu and Collier~\cite{hu2024persona} found that persona prompting produced modest improvements when persona variables were relevant to subjective judgments, but its benefits were limited when those variables explained little human variation. Agnew et al.~\cite{agnew2024inclusion} further caution that replacing human participants with AI surrogates does not provide the representation, inclusion, or lived experience of actual participation. Accordingly, we use personas as diagnostic probes rather than substitutes for human audiences or estimates of population prevalence. We compare the sampled takeaways to identify potential ambiguity in a visualization, not to predict how actual audiences will interpret it in practice.

\section{Problem Formulation and Methodology} \label{sec:methodology}

Interpretive ambiguity becomes observable when the same visualization supports meaningfully different chart-grounded takeaways. We study this variation by eliciting multiple contextual takeaways for the same interpretive question, extracting their statements about perceived visual patterns, and assigning ambiguity labels to pairs of these statements across personas. The distribution of ambiguity labels forms an \textit{ambiguity profile} that surfaces potential ambiguity for an author to investigate.

This section defines our analytical units, motivates persona-conditioned sampling, and presents the four-stage ambiguity-profiling pipeline. \autoref{sec:tech-eval} describes its chart corpus and VLM selection, and \autoref{sec:amb-framework} derives and operationalizes the ambiguity taxonomy.

\begin{figure*}[t]
    \centering
    \includegraphics[width=\textwidth]{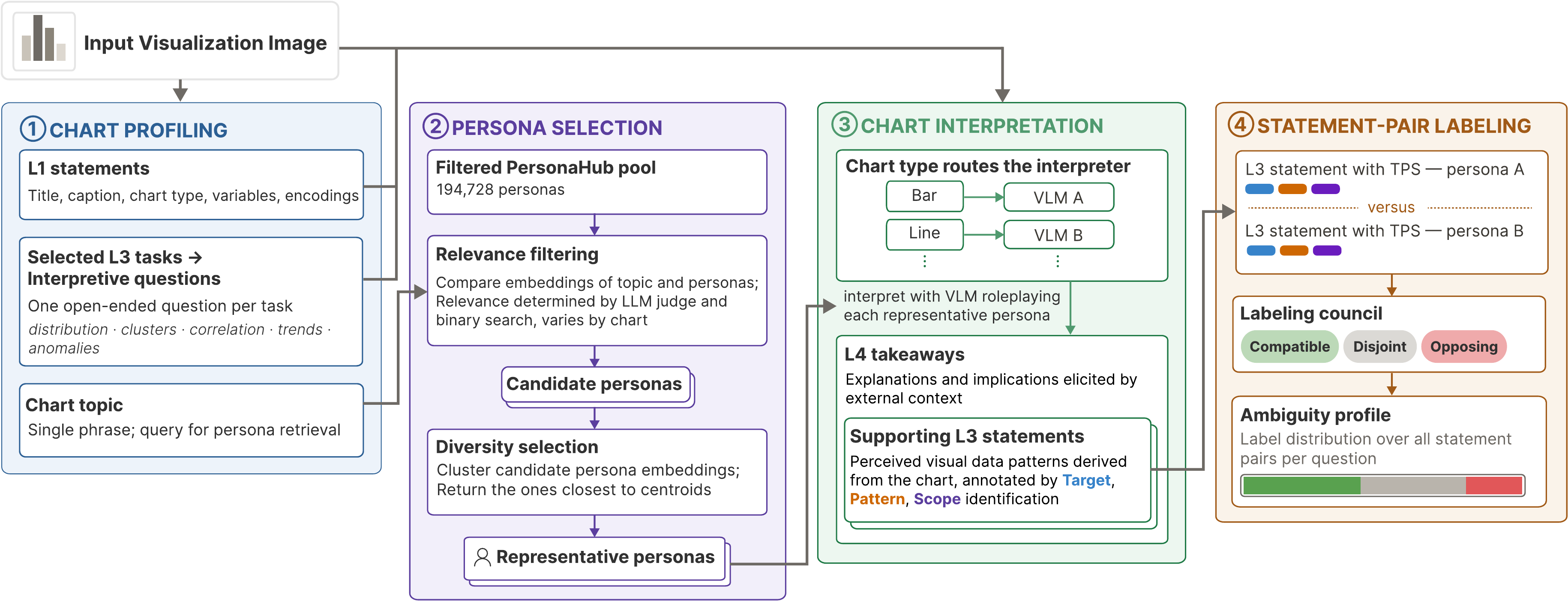}
    \caption{Overview of the four-stage ambiguity-profiling pipeline. Given a visualization, (1) chart profiling constructs a chart profile containing an L1 statement, topic, selected L3 tasks, and interpretive questions; (2) persona selection retrieves relevant and diverse representative personas; (3) a chart-type-routed VLM generates contextual takeaways (L4) and supporting L3 statements decomposed into Target, Pattern, and Scope; and (4) a labeling council assigns Compatible, Disjoint, or Opposing ambiguity labels to statement pairs and aggregates them into an ambiguity profile. The workflow is model-agnostic, although our implementation routes each chart type to its selected VLM. }
    \label{fig:pipeline}
\end{figure*}
\subsection{Operationalizing Interpretive Ambiguity} \label{sec:framework-foundation}

\textbf{Definition.}
We define interpretive ambiguity as the potential for a visualization to support substantively different claims about its visible patterns in response to the same interpretive question. Such claims foreground different visual evidence or make distinct or conflicting assertions rather than restating the same idea.

\textbf{Analytical scope and units.}
Lundgard and Satyanarayan's four-level model distinguishes chart construction and local values (L1 and L2), visible patterns (L3), and takeaways connected to external context (L4) \cite{lundgard2021accessible}. We analyze \textit{L3 statements}: chart-grounded claims about patterns, including extrema, outliers, correlations, and comparisons. An \textit{L4 takeaway} is a persona's complete contextual takeaway in response to an interpretive question and may contain multiple supporting L3 statements. A \textit{statement pair} contains two L3 statements produced by different personas for the same chart and question. The question-specific distribution of ambiguity labels across statement pairs constitutes an \textit{ambiguity profile}.

\textbf{Boundaries.}
We compare only L3 statements intended to be supported by visible chart evidence. Their ambiguity labels characterize semantic relationships rather than factual correctness, which makes chart comprehension and grounding prerequisites. Ambiguity profiles are diagnostic: they neither estimate interpretation prevalence nor declare a chart definitively ambiguous.

\textbf{Initial ambiguity-label vocabulary.}
Natural Language Inference (NLI) provides a starting point for comparing whether meanings follow from, contradict, or remain neutral to one another \cite{poliak2020survey}. Because our pairs have no fixed premise--hypothesis direction, we began with four working ambiguity labels: Equivalent statements express the same claim; Entailment indicates that one statement implies the other; Disjoint statements describe different, jointly acceptable aspects; and Opposing statements conflict directly or through framing.

We renamed NLI's Neutral relation as \hlD{Disjoint} to emphasize shared grounding in one visualization and broadened Contradiction to \hlO{Opposing} to include conflicting higher-level framing. Qualitative analysis later consolidated Equivalent and Entailment as \hlN{Compatible} (\autoref{sec:label-perf-review}).

\subsection{Persona-based VLM Takeaway Sampling} \label{sec:persp-sim}

Eliciting varied interpretations from human participants is costly and difficult to scale \cite{borgo2018information}. We therefore use VLM role-playing to probe plausible variation rather than estimate its prevalence in a population \cite{shanahan2023role,salewski2023impersonation}.

Our fixed pool contains 194,728 filtered descriptions from PersonaHub \cite{personahub}. Entries typically combine a social or professional role with interests, concerns, goals, or attitudes; for example, \textit{``A fellow government official who is unsure about the impact of stricter regulations on oil drilling and seeks more information.''} Filtering details are provided in the supplemental material.

\subsection{Ambiguity Profiling Pipeline} \label{sec:pipeline-design}

The ambiguity-profiling pipeline in \autoref{fig:pipeline} has four stages: (1) chart profiling, (2) persona selection, (3) chart interpretation, and (4) statement-pair labeling. Chart profiling is performed once per chart; persona selection uses the extracted topic, chart interpretation combines the selected personas with each interpretive question, and statement-pair labeling compares the resulting L3 statements within each question.

\subsubsection{Chart Profiling} \label{sec:chart-profiling}

Chart profiling transforms the input image into an L1 statement, a chart topic, selected tasks, and interpretive questions shared by later stages.

\textbf{Chart description and topic extraction.}
An image-grounded VLM describes the chart's title or caption, type, axes, variables, encodings, and categories. This L1 statement provides shared chart information and the type used for model routing. Using the complete chart rather than an isolated title or caption, the VLM also summarizes its subject in a single phrase that serves as the query for persona retrieval.

\textbf{Task selection and question generation.}
The VLM selects applicable L3 tasks---characterizing distributions, identifying clusters, examining correlations, describing large and small trends, and finding anomalies \cite{lee2006task,stokes2025writerankratecomparing}---and generates one open-ended question per task. Tasks are selected from the visible structure of each chart rather than assigned solely by chart type, allowing charts of the same type to support different analyses. Unlike ChartQAPro's largely factual questions, these questions permit multiple chart-grounded responses. Asking every persona the same question controls the subsequent comparison.

\textbf{Implementation.}
In our implementation, GPT 5.6 Terra produces the complete chart profile once in a single image-grounded call. Because chart-type-specific model routing depends on the type recorded in the L1 statement, routing begins only at chart interpretation; full task definitions and prompts appear in the supplemental material.

\subsubsection{Persona Selection} \label{sec:persona-select}
Persona selection takes the chart topic and filtered PersonaHub pool as input and returns a set of relevant but varied representative personas.

\textbf{Relevance filtering.}
The stage embeds the chart topic and persona descriptions, ranks personas by cosine similarity, and uses an LLM judge with binary search to set a chart-specific relevance threshold. This judge-guided search tests personas near the candidate boundary rather than evaluating the entire pool.

\textbf{Diversity selection.}
Relevance filtering alone may return many semantically similar personas. The stage therefore clusters the candidate embeddings to identify distinct regions of the relevant pool. One estimate captures separation in the full candidate set, while another measures how many groups remain stable under subsampling. We use the smaller estimate to avoid selecting more representatives than either criterion supports, then return each cluster's medoid as a representative persona. Embedding-space diversity does not guarantee demographic or ideological coverage, so the representatives are not an audience sample.

\textbf{Implementation.}
OpenAI's \texttt{text-embedding-3-small} embeds and ranks the candidates, while GPT 5.6 Terra judges their relevance. Dimensionality reduction, clustering parameters, stopping rules, and pilot results appear in the supplemental material.

\subsubsection{Chart Interpretation} \label{sec:chart-interp}

Chart interpretation takes the chart image, its L1 statement, one interpretive question, and one persona, and returns an L4 takeaway with supporting chart-grounded L3 statements.

\textbf{Takeaway generation.}
A VLM role-plays the persona and produces an L4 takeaway reflecting what that simulated viewer might foreground or emphasize. The persona conditions the selection and characterization of visible patterns, while the chart remains the required evidential basis.

\textbf{Grounding and decomposition.}
The same VLM identifies the visible observations and L3 statements supporting the takeaway, then decomposes each L3 statement into Target, Pattern, and Scope for statement-pair labeling. TPS is developed in \autoref{sec:tps} and operationalized in \autoref{sec:pipeline-opt}.

\textbf{Implementation.}
Each persona--question pairing uses one image-grounded call to the chart-type-specific model selected in \autoref{sec:tech-eval}, with a prompt adapted from Charts-of-Thought \cite{das2025charts}. The model generates the L4 takeaway before its L3 support because preliminary experiments found that this ordering produced L3 statements that were more responsive to the question and less likely to introduce information not visible in the chart. It extracts TPS in the same call; pilot results and prompt details appear in the supplemental material.

\begin{table*}[h]
    \centering
    \begin{tabular}{|l|c|c|c|c|c|c|c|c|c|c|}
    \hline
        \multicolumn{2}{|l|}{} & \multicolumn{3}{c|}{Fast Tier} & \multicolumn{3}{c|}{Balanced Tier} & \multicolumn{3}{c|}{Performance Tier} \\
    \hline
          & & \multirow{3}{3em}{GPT 5.6 Luna} & \multirow{3}{4em}{Gemini 3.5 Flash-Lite} & \multirow{3}{3em}{Claude Haiku 4.5} & \multirow{3}{3em}{GPT 5.6 Terra} & \multirow{3}{4em}{Gemini 3.6 Flash} & \multirow{3}{3em}{Claude Sonnet 5} & \multirow{3}{3em}{GPT 5.6 Sol} & \multirow{3}{4em}{Gemini 3.1 Pro Preview} & \multirow{3}{3em}{Claude Opus 5}\\
          \textbf{Chart Type} & \textbf{N} & & & & & & & & & \\
          & & & & & & & & & & \\          
        \hline
        Line              & 231 & 72.73\% & 64.94\% & 51.52\% & 70.56\% & \textbf{80.52\%} & 71.00\% & 78.35\% & 77.92\% & 77.49\%\\
        Bar               & 212 & 85.85\% & 72.64\% & 58.49\% & 84.43\% & \textbf{87.74\%} & 85.38\% & 87.26\% & 86.79\% & 85.85\%\\
        Stacked Bar       & 94  & 72.34\% & 61.70\% & 61.70\% & 73.40\% & \textbf{84.04\%} & 74.47\% & 76.60\% & 75.53\% & 70.21\%\\
        Pie               & 20  & 95.00\% & 80.00\% & 55.00\% & 90.00\% & 95.00\% & 95.00\% & \textbf{100.00\%}& \textbf{100.00\%} & \textbf{100.00\%} \\
        Scatterplot       & 17  & 70.59\% & 70.59\% & 47.06\% & 64.71\% & 76.47\% & 82.35\% & 70.59\% & 82.35\% & \textbf{88.24\%} \\
        Area              & 26  & 73.08\% & 69.23\% & 61.54\% & 65.38\% & 73.08\% & 65.38\% & 76.92\% & \textbf{80.77\%} & \textbf{80.77\%} \\
        Stacked Area      & 14  & \textbf{78.57\%} & 57.14\% & 42.86\% & 64.29\% & 71.43\% & 50.00\% & 57.14\% & 71.43\% & 64.29\% \\
        Bar + Line        & 75  & 76.00\% & 57.33\% & 52.00\% & 69.33\% & 81.33\% & 74.67\% & 73.33\% & \textbf{85.33\%} & 76.00\% \\
        \hline
        Overall           & 689 & 77.79\% & 66.62\% & 55.30\% & 75.18\% & \textbf{83.16\%} & 76.63\% & 80.26\% & 81.86\% & 79.68\%\\
        \hline
        \multicolumn{2}{|l|}{Cost (USD)} & \$1.34 & \$0.92 & \$3.60 & \$12.79 & \$11.79 & \$8.54 & \$40.26 & \$26.29 & \$23.31 \\
        \hline
    \end{tabular}
    \caption{ChartQAPro accuracy by chart type. Answers were scored by a Gemini 3.5 Flash-Lite judge; cost uses provider prices from August 2026.}
    \label{tab:model-selection}
\end{table*}
\subsubsection{Statement Pair Construction and Labeling} \label{sec:pairwise-labeling}

The final stage assigns ambiguity labels to pairs of generated L3 statements and aggregates the labels into a question-specific ambiguity profile.

\textbf{Pair construction.}
For each interpretive question, the stage pairs every combination of L3 statements produced by different personas. Restricting comparisons to different personas answering the same question controls for the interpretive prompt, so the ambiguity labels capture variation among persona-conditioned responses. The number of comparisons grows quadratically with the number of generated L3 statements.

\textbf{Ambiguity labeling and profile construction.}
Language-model judges use the chart's L1 statement and each pair's TPS components to assign a \hlN{Compatible}, \hlD{Disjoint}, or \hlO{Opposing} ambiguity label from \autoref{sec:taxonomy}. The normalized distribution of ambiguity labels across all statement pairs for the interpretive question constitutes its ambiguity profile.

\textbf{Implementation.}
In our implementation, labeling is text-only: judges receive the L1 statement and TPS components rather than the chart image. The supplemental material gives the pair-count expression, and \autoref{sec:pipeline-opt} describes and evaluates the labeling council.

\section{Dataset Curation and VLM Selection} \label{sec:tech-eval}

Instantiating the pipeline requires two technical choices: charts on which to develop and apply it, and VLMs capable of reliably reading the included chart types. We first curate 783 charts and 689 chart--question pairs, then use the pairs to benchmark candidate VLMs and select a model for generating persona-conditioned L4 takeaways for each chart type. The benchmark evaluates chart comprehension, a prerequisite for generating grounded L4 takeaways.

\subsection{Dataset Curation} \label{sec:dataset-selection}

We curated a dataset with two core components: a set of unique charts for developing the pipeline and a set of chart--question pairs for selecting VLMs.

\textbf{Dataset selection.}
We sought real-world, public-facing charts spanning the VLAT chart types and paired with questions that test chart comprehension \cite{vlat}. Among the visualization-literacy datasets we reviewed \cite{wang2024charxiv,masry2022chartqa,masry2025chartqapro,zheng2025advancing,roberts2024scifi,xie2025infochartqa,li2025chartgalaxy,huang2025evochart}, ChartQAPro \cite{masry2025chartqapro} offered the best combination of relevant chart types and question coverage. We did not verify every chart against design guidelines, so the corpus is not a certified set of well-designed visualizations; detailed selection criteria appear in the supplemental material.

\textbf{Chart curation.}
Exact deduplication reduced ChartQAPro's 1,948 chart--question pairs to 1,252 unique charts. Because ChartQAPro lacks chart-type metadata, a three-VLM council assigned types by majority vote and researchers resolved the remaining 52 charts. We merged structurally similar sparse types, excluded out-of-scope and underrepresented categories, and supplemented scarce pie and scatterplot charts from ChartQA \cite{masry2022chartqa} and ChartMuseum \cite{tang2026chartmuseum}. The resulting chart corpus contains 783 unique charts; council configuration, category mappings, exclusions, and source counts appear in the supplemental material.

\textbf{Chart--question pair curation.}
For the VLM benchmark, we retained answerable ChartQAPro factoid, multiple-choice, and fact-checking questions associated with the final chart types, excluding conversational and hypothetical questions. This produced 689 chart--question pairs, summarized with the chart corpus in \autoref{tab:chart-type-count-dataset}.

\begin{table}[h]
    \centering
    \begin{tabular}{|l|c|c|}
        \hline
        \textbf{Chart Type} & \textbf{Charts} & \textbf{Chart--Question Pairs} \\
        \hline
        Line                  & 263 & 231 \\
        Bar                   & 194 & 212 \\
        Stacked Bar           & 98  & 94 \\
        Pie                   & 44  & 20 \\
        Scatterplot           & 48  & 17 \\
        Area                  & 31  & 26 \\
        Stacked Area          & 17  & 14 \\
        Bar + Line            & 88  & 75 \\
        \hline
        \textbf{In-scope total} & \textbf{783} & \textbf{689} \\
        \hline
    \end{tabular}
    \caption{Chart-type distributions of the curated dataset's two components. Chart counts are unique images and include supplemental charts; counts of chart--question pairs include retained ChartQAPro questions.}
    \label{tab:chart-type-count-dataset}
\end{table}

\subsection{VLM Selection} \label{sec:model-selection}

Before eliciting persona-conditioned L4 takeaways, the pipeline needs a VLM that can reliably read each chart. Because chart-comprehension performance can vary across chart types, we benchmarked candidate VLMs and routed each type to its highest-scoring model, reducing chart-reading errors as a source of variation in the resulting takeaways.

\subsubsection{Candidate Models}

Although small, chart-specialized open-source models can match much larger proprietary VLMs on chart QA \cite{dong2026attn}, our pipeline requires general-purpose models that support open-ended, persona-conditioned, structured generation through hosted APIs. We therefore benchmarked nine proprietary VLMs spanning the GPT, Gemini, and Claude families and provider-described fast, balanced, and performance tiers in August 2026 (\autoref{tab:model-selection}). Model-selection and pricing details appear in the supplemental material.
We selected one model from each provider at each tier to cover variation both across model families and within each family. The tier labels organize the candidate set rather than define analytical categories.

\begin{table*}[t]
    \centering
    \small
    \begin{tabular}{|p{0.13\textwidth}|p{0.20\textwidth}|p{0.57\textwidth}|}
        \hline
        \textbf{Label} & \textbf{Subtype} & \textbf{Relationship among TPS components} \\
        \hline
        Compatible & \textbf{Equivalent (EQ)} & Target, Pattern, and Scope are equivalent in chart context. \\
        Compatible & \textbf{Directional entailment} & One or more TPS components are more specific in one consistent direction; no component creates conflict. \\
        \hline
        Disjoint & \textbf{Complementary parts (D1)} & Targets or Scopes partition a shared whole, but neither statement entails the other. \\
        Disjoint & \textbf{Target separation (D2)} & Targets are non-overlapping sibling entities or categories. \\
        Disjoint & \textbf{Scope separation (D3)} & Scopes cover non-overlapping ranges or conditions. \\
        Disjoint & \textbf{Pattern separation (D4)} & Patterns describe fundamentally different properties without creating conflict. \\
        Disjoint & \textbf{Mixed TPS (D5)} & Specificity relations across TPS components point in different entailment directions. \\
        \hline
        Opposing & \textbf{Direct contradiction (O1)} & Related Targets and Scopes support directly incompatible Patterns or claims. \\
        Opposing & \textbf{Conflicting framing (O2)} & Related evidence is characterized through Patterns or emphasis that create conflicting overall impressions. \\
        \hline
    \end{tabular}

    \caption{Hierarchy of top-level ambiguity labels and their explanatory subtypes.}
    \label{tab:taxonomy-summary}
\end{table*}

\subsubsection{Benchmark Procedure} \label{sec:model-eval-setup}

We evaluated all nine models on the 689 chart--question pairs using identical response-generation and scoring procedures.

\textbf{Model response generation.}
Each model received the chart image, question, and the same Charts-of-Thought-based prompt for extracting, checking, and reasoning over chart values \cite{das2025charts}. Responses used a shared structured format; configuration and retry rules appear in the supplemental material.

\textbf{Response scoring.}
Gemini 3.5 Flash-Lite judged anonymized responses against ChartQAPro reference answers using semantic rather than exact-match scoring. This accommodated equivalent answer formats, such as a percentage expressed as 4\%, 4, 0.04, or within a complete sentence. Answers involving years received a one-year tolerance, and other numerical answers received a 5\% tolerance. Calls that failed after all retries and unintelligible or malformed responses were counted as incorrect. Overall accuracy was computed over all 689 pairs; additional judge settings appear in the supplemental material.

\subsubsection{Results and Model Routing}

\textbf{Benchmark results and routing.}
Higher provider-described tiers did not consistently improve accuracy, and no single model performed best across all chart types (\autoref{tab:model-selection}). Gemini 3.6 Flash achieved the highest overall accuracy (83.16\%) and led the three largest categories---line, bar, and stacked bar---but models from other tiers and families performed best on the remaining five types. This variation motivates chart-type-specific routing rather than using the overall winner for every chart.

During chart interpretation, the pipeline uses the chart type recorded in the chart profile to select the highest-scoring model for that type: Gemini 3.6 Flash for line, bar, and stacked bar; Gemini 3.1 Pro Preview for pie, area, and bar + line; Claude Opus 5 for scatterplots; and GPT 5.6 Luna for stacked area. Ties are resolved using overall benchmark accuracy. For chart types absent from the benchmark, the pipeline defaults to Gemini 3.6 Flash, the model with the highest overall accuracy. Complete implementation details appear in the supplemental material.

\section{The Ambiguity Framework} \label{sec:amb-framework}

The preceding sections defined the ambiguity-profiling pipeline and established the chart corpus and routed VLMs used to generate L4 takeaways. Preliminary experiments showed that the VLM's assignments under the four NLI-inspired ambiguity labels in \autoref{sec:framework-foundation} sometimes differed from our judgments and changed across repeated calls. These inconsistencies raised the possibility that the problem lay not only in model performance, but also in how the ambiguity labels were conceptualized. We therefore conducted a qualitative review to identify the sources of disagreement and refine the ambiguity framework. This section first describes that process, then presents Target, Pattern, and Scope (TPS) and our final framework of \hlN{Compatible}, \hlD{Disjoint}, and \hlO{Opposing}, and finally operationalizes and evaluates the framework as the pipeline's statement-pair labeling stage.

\subsection{Qualitative Framework-Derivation Procedure} \label{sec:label-perf-review}

The qualitative review comprised generating and sampling prototype statement pairs, independently applying the initial ambiguity labels, and refining the framework through negotiated agreement.

\textbf{Prototype generation and sampling.} We ran a prototype of the pipeline on eight charts, one selected from each of the chart types retained in \autoref{sec:tech-eval}. This sample provided structurally varied cases for examining how the initial ambiguity labels behaved across chart types. At this stage, statement-pair labeling used the four working ambiguity labels---Equivalent, Entailment, Disjoint, and Opposing---rather than the final three-label taxonomy developed below. From the resulting output, we sampled 115 statement pairs for qualitative analysis. We sought approximately balanced coverage across chart types and VLM-assigned ambiguity labels so that the analysis would not be dominated by common chart types or ambiguity-label categories.

\textbf{Independent labeling and negotiated agreement.} Three authors independently labeled each pair using the four initial definitions without seeing the VLM's decision. They then discussed every non-unanimous ambiguity label and used recurring sources of disagreement to refine the distinctions and their definitions. As the framework developed, they applied it across the full sample and assigned the components described in \autoref{sec:tps}. The process concluded when all three authors agreed on the definitions and on the ambiguity label and component assignments for every pair, producing 115 human-authored reference annotations.

\subsection{Target, Pattern, Scope} \label{sec:tps}
Negotiated agreement revealed that distinctions between ambiguity labels repeatedly turned on three core dimensions. TPS captures the aspects of an L3 statement needed to determine whether it expresses the same, a different, or a conflicting reading relative to another statement. We represent these aspects as a \tpsPattern{\textbf{Pattern}} asserted of one or more \tpsTarget{\textbf{Targets}} within a \tpsScope{\textbf{Scope}}. Target identifies the visible data entities being discussed, Pattern specifies what is claimed about them, and Scope bounds where, when, or under which conditions the claim applies. This structure parallels predicate--argument representations in linguistics, which distinguish a predicate from its participants and modifiers \cite{palmer2005propbank}, as well as visualization data-fact models that distinguish what is characterized from the data entities and subspace involved \cite{wang2020datashot}.

The colors below identify the corresponding components in the L3 statement examples throughout this section:

\vspace{0.5em}
\textit{The \tpsTarget{president's approval rating among all voters} \tpsPattern{generally declined} \tpsScope{from February 2021 to July 2022}.}
\vspace{0.5em}

Each TPS component can be expressed at different levels of specificity. A Target becomes more specific by referring to a subset of entities, a Pattern by strengthening or quantifying the same underlying relationship, and a Scope by narrowing the applicable range or conditions. These relationships are evaluated in chart context. The following L3 statement is more specific along all three dimensions:

\vspace{0.5em}
\textit{The \tpsTarget{president's approval rating among voters aged 18--29} \tpsPattern{declined by 10 percentage points} \tpsScope{from February to July 2021}.}
\vspace{0.5em}

Not every L3 statement expresses all three components explicitly. For example, a comparison may identify multiple Targets and a Pattern without specifying a meaningful Scope.

Together, these components provide a common basis for identifying where two L3 statements align, diverge, or conflict, and thereby for defining the \hlN{Compatible}, \hlD{Disjoint}, and \hlO{Opposing} ambiguity labels introduced in the next section. Each component may contain multiple elements, remain implicit, and vary in specificity, allowing TPS to represent comparisons, compound patterns, and conditional claims without adding further component types. Across the 115 statement pairs examined during negotiated agreement, every distinction that affected ambiguity labeling could be captured through TPS or interactions among its components; no additional recurring component emerged. We therefore consider TPS expressively sufficient for the chart-grounded L3 statements in our analysis, while making no corresponding claim about arbitrary natural-language discourse.

\subsection{Ambiguity Taxonomy} \label{sec:taxonomy}

\begin{figure*}[t]
    \centering
    \includegraphics[width=\textwidth]{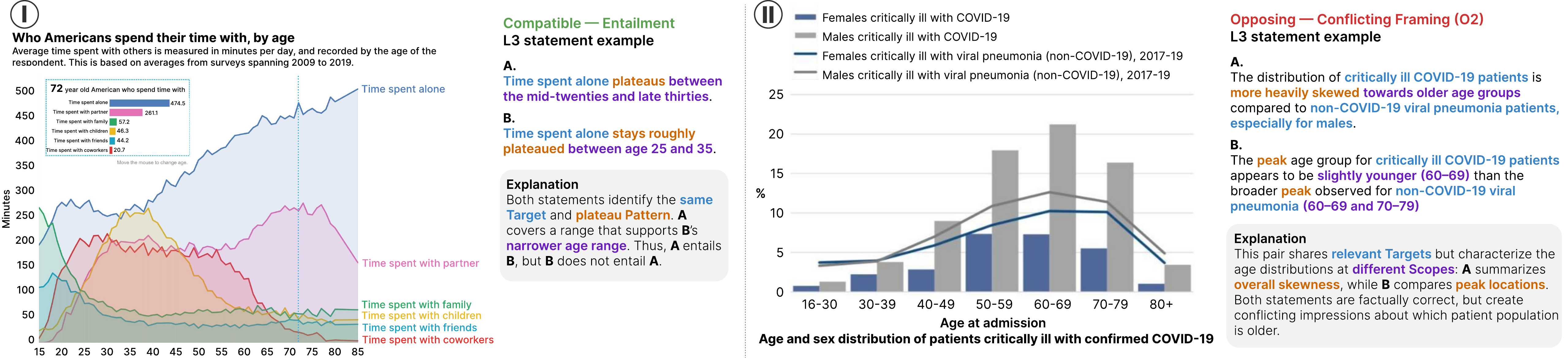}
    \caption{Examples of L3 statement pairs and their ambiguity labels. (I) Two statements identify the same Target and Pattern, with statement A covering an age range that includes the Scope in statement B, leading to a Compatible label through entailment. (II) Two factually sustainable statements emphasize different Patterns at different Scopes, leading to an Opposing label through conflicting framing. }
    \label{fig:example-charts}
\end{figure*}


The second outcome of our qualitative analysis was the consolidation of the four initial ambiguity labels into three top-level ambiguity labels. Because Equivalent and Entailment both preserve a shared reading, we group them under \hlN{Compatible} while retaining equivalence and directional entailment as subtypes.

The taxonomy compares statement pairs at two levels: the top-level ambiguity label characterizes whether their complete readings are shared, different, or conflicting, while the subtype identifies the TPS configuration responsible for that label. \hlO{Opposing} takes precedence when the statements contradict one another or create conflicting impressions. Otherwise, equivalence or one-way entailment produces \hlN{Compatible}, while pairs with neither entailment nor conflict are \hlD{Disjoint}. \autoref{tab:taxonomy-summary} summarizes the hierarchy defined below, which we illustrate using statement pairs grounded in the two charts in \autoref{fig:example-charts}.

\subsubsection{Compatible}

A pair receives the \hlN{Compatible} ambiguity label when its statements express the same reading or establish entailment in one direction without conflict. \textbf{Equivalent (EQ)} applies when all TPS components are equivalent; \textbf{directional entailment} applies when one or more components are more specific in one consistent direction. The specificity comparison in \autoref{sec:tps} illustrates how TPS establishes this direction; an additional chart-grounded entailment pair appears in the supplemental material.

\subsubsection{Disjoint}

A pair receives the \hlD{Disjoint} ambiguity label when TPS differences prevent entailment without creating conflict. \autoref{tab:taxonomy-summary} identifies the source of separation captured by each subtype; pairs fitting none retain only the top-level ambiguity label. The following pair from the time-use chart in \autoref{fig:example-charts}-I illustrates Pattern separation (\textbf{D4}):
\vspace{0.5em}

\noindent \textbf{A:} \textit{\tpsScope{The period between ages 25 and 45} \tpsPattern{shows the lowest levels of} \tpsTarget{time spent alone}.}
\vspace{0.3em}

\noindent \textbf{B:} \textit{\tpsTarget{The amount of time spent alone} \tpsPattern{generally increases with age}, \tpsScope{especially after the mid-20s}.}
\vspace{0.5em}

The statements share a Target and overlapping Scopes but describe a nonconflicting extremum and trend, producing the \hlD{Disjoint} ambiguity label through Pattern separation.

\subsubsection{Opposing}

A pair receives the \hlO{Opposing} ambiguity label when related evidence supports directly incompatible claims or conflicting overall impressions. \textbf{Direct contradiction (O1)} applies when the statements cannot both hold as written. For the time-use chart in \autoref{fig:example-charts}-I:
\vspace{0.5em}

\noindent \textbf{A:} \textit{\tpsTarget{Time spent alone} \tpsPattern{initially decreases} \tpsScope{from age 15 to 20}, then \tpsPattern{steadily rises} \tpsScope{throughout adulthood and into older age}.}
\vspace{0.3em}

\noindent \textbf{B:} \textit{\tpsTarget{Time spent alone} \tpsPattern{initially decreases slightly} \tpsScope{after age 20} before beginning \tpsPattern{a steady upward trend}.}
\vspace{0.5em}

The shared Target and overlapping Scopes place the change in direction at incompatible points, producing the \hlO{Opposing} ambiguity label through direct contradiction.

\textbf{Conflicting framing (O2)} applies when both statements are factually sustainable, but their framing or emphasis creates conflicting overall impressions. As shown in~\autoref{fig:example-charts}-II,
statement \textbf{A} summarizes overall distributions, while \textbf{B} compares their peak locations. Both statements hold, but their Target--Pattern compositions create conflicting impressions of which patient population is older, producing the \hlO{Opposing} ambiguity label through conflicting framing rather than Pattern separation.

\subsection{Operationalizing and Evaluating the Taxonomy} \label{sec:pipeline-opt}

The taxonomy provides conceptual definitions for assigning ambiguity labels to pairs of L3 statements, but applying those definitions across many generated statements requires translating them into a computational pipeline. Within the ambiguity-profiling pipeline described in \autoref{sec:pipeline-design}, this translation involves two connected operations: TPS identification produces a structured representation of each L3 statement during chart interpretation, and statement-pair labeling uses those representations to assign ambiguity labels.

\textbf{TPS identification.} During chart interpretation, the selected VLM produces a TPS description alongside each generated L3 statement. We considered sentence partitioning, which marks contiguous spans but cannot represent implicit or distributed components, and in-context referral, which asks the image-grounded VLM to describe each component. We selected in-context referral because it accommodates the varied component structures in \autoref{sec:tps} and resolves implicit references using the chart without requiring a separate model call.

\textbf{Statement-pair labeling.} After chart interpretation, the pipeline constructs every within-question, between-persona statement pair. For each pair, the statement-pair labeling stage receives the chart's L1 statement, the two L3 statements, and their TPS components. It returns a \hlN{Compatible}, \hlD{Disjoint}, or \hlO{Opposing} ambiguity label, the corresponding subtype when applicable, and an explanation grounded in the pair's TPS components. For directional entailment, it also identifies which statement entails the other.

We considered two implementations of this stage. The gated decision tree asks a sequence of questions about TPS overlap, specificity, implication, and conflict, which code maps to an ambiguity label. The holistic strategy instead considers the complete statements, assesses all three ambiguity labels in parallel, and returns an explanation, subtype, and entailment direction when applicable.

\textbf{Selecting the statement-pair labeling strategy.} From the 115 human-annotated pairs, we selected 40 for prompt development using stratified random sampling to balance chart types and ambiguity labels; the remaining 75 formed the evaluation set. Gemini 3.5 Flash-Lite performed the statement-pair labeling, while Claude Sonnet 5 assisted with prompt revision. We scored both partitions at each iteration but revised the prompts using only development-set examples and errors.

The holistic strategy achieved higher exact agreement with the negotiated top-level ambiguity labels on both partitions (\autoref{tab:labeling-methods}), so we adopted it for the final pipeline. This evaluation concerns only the three top-level ambiguity labels; TPS identification and subtype assignment were not evaluated.

\begin{table}[h]
    \centering
    \begin{tabular}{|l|c|c|}
        \hline
        \textbf{Labeling Strategy} & \textbf{Development} & \textbf{Evaluation} \\
        \hline
        Gated Decision Tree &  33/40 (82.50\%) & 52/75 (69.33\%)\\
        Holistic & 37/40 (92.50\%) & 54/75 (72.00\%) \\
        \hline
    \end{tabular}
    \caption{Exact agreement with negotiated top-level ambiguity labels for two implementations of the statement-pair labeling stage.}
    \label{tab:labeling-methods}
\end{table}

\textbf{Final pipeline configuration.} The final statement-pair labeling stage applies the holistic strategy through a council comprising Gemini 3.6 Flash, GPT 5.6 Terra, and Claude Sonnet 5. These balanced-tier models represent three provider families, reducing dependence on one model family without incurring the cost of each provider's largest model.

Each council member receives one statement pair, the chart's L1 statement, and the respective Target, Pattern, and Scope of both L3 statements. We use one text-only prompt per pair because batched prompts sometimes omitted comparisons and image-based calls would substantially increase cost. Each model returns a top-level ambiguity label, explanation, and subtype when applicable. The council votes only on the top-level ambiguity label; subtype outputs are retained for subsequent analysis but are neither reconciled nor evaluated for consistency. A three-way split is retained as Divided. Across all 115 human-labeled pairs, the final council classified 101 correctly (87.8\%); two pairs were Divided. This aggregate result evaluates the final council, whereas \autoref{tab:labeling-methods} reports the held-out evaluation used to select the statement-pair labeling strategy.

\textbf{Pipeline output: ambiguity profiles.} The pipeline aggregates the council's ambiguity labels into question-specific ambiguity profiles and reports the four outcomes as normalized proportions, enabling comparison despite varying statement-pair counts. Model-generated subtype annotations and TPS explanations are retained to explain individual labels. Across the complete output, 71 of 11,083 statement pairs (0.64\%) were Divided. The case studies in the following section interpret these distributions as descriptive signals alongside their underlying subtype annotations, TPS descriptions, and L4 takeaways.

\section{Case Studies}

We apply our methodology to 24 real-world, public-facing visualizations curated from our pipeline-development dataset and EvoChart~\cite{huang2025evochart}. 
The collection is balanced across chart types and covers socially consequential topics that we expect to support distinct perspectives. 
We first summarize the resulting ambiguity profiles and then use two cases to examine how L3 statements relate to L4 takeaways and how a visualization redesign changes chart interpretations.

\begin{figure*}[t]
    \centering
    \includegraphics[width=\textwidth]{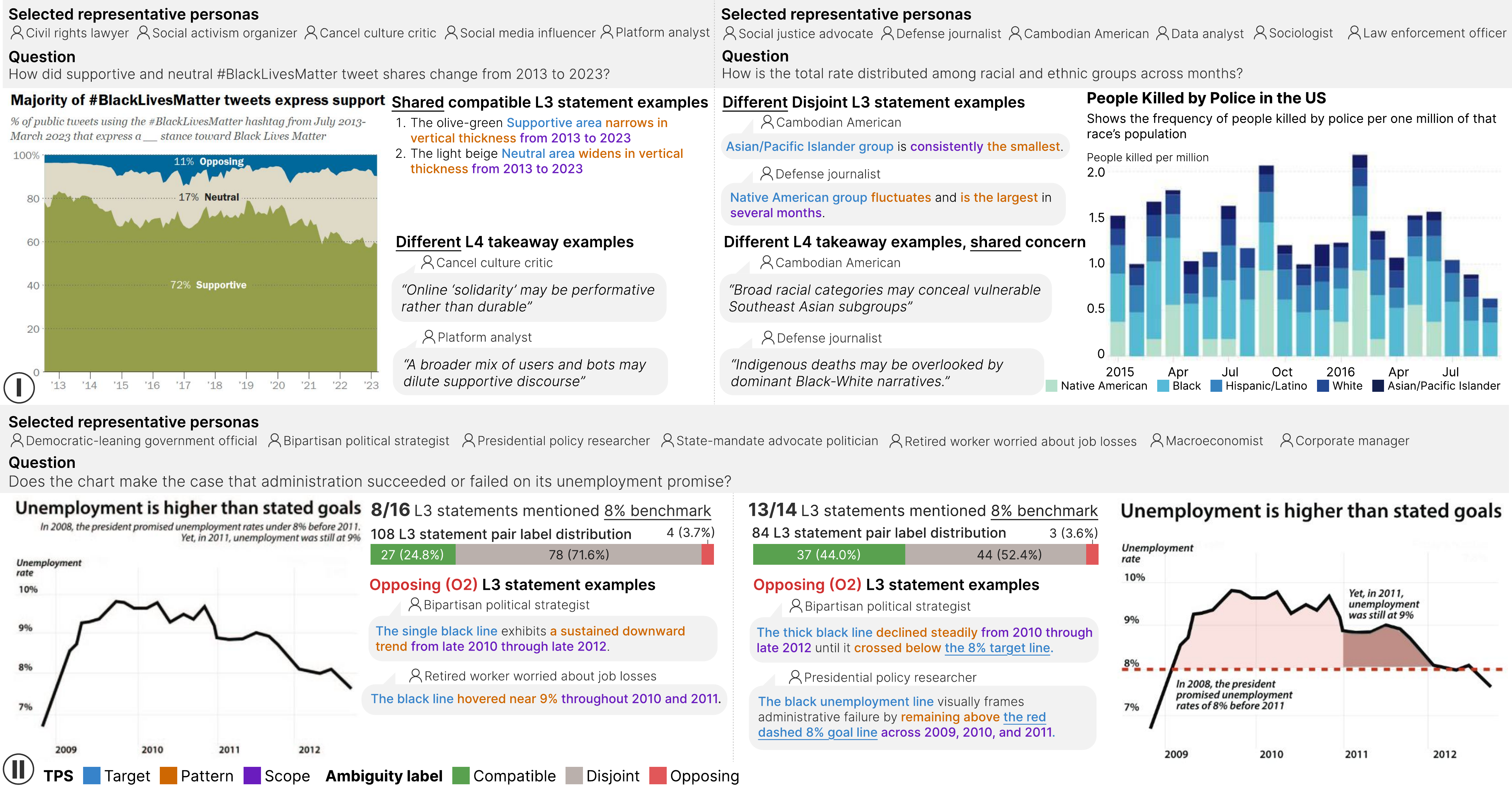}
    \caption{Case studies illustrating relationships between L3 statements and L4 takeaways, and changes following a visualization redesign. (I) For the \#BlackLivesMatter chart, Compatible L3 statement pairs identify the same decline in supportive sentiment but support different L4 takeaways across personas. For the police violence chart, Disjoint L3 statement pairs focus on different racial groups and patterns, while their L4 takeaways share a broader concern about obscured racial disparities. (II) The original and redesigned unemployment charts use the same data, personas, and question. The redesign more frequently anchors L3 statements to the 8\% benchmark and shifts the ambiguity-label distribution from Disjoint toward Compatible; representative Opposing pairs nevertheless emphasize either subsequent recovery or prolonged failure.}
    \label{fig:case_study}
\end{figure*}

\textbf{Overall findings.} 
Across the 24 visualizations, our pipeline generates 66 interpretive questions, 166 selected personas, 1,504 L3 statements, and 11,083 statement pairs.
On average, it is 2.75 interpretive questions per chart, 6.92 personas per chart, and 3.23 L3 statements per persona per question.
Of these L3 statement pairs, 2,945 (26.6\%) were Compatible, 7,655 (69.1\%)  Disjoint, and 412 (3.7\% ) Opposing, with the remaining 71 Divided.
The predominance of Disjoint labels is expected because we compare all cross-persona statement pairs, many of which concern different Targets, Scopes, or Pattern types, satisfying \textbf{D2}, \textbf{D3}, or \textbf{D4}.
Among the 412 Opposing pairs, 178 (43.2\%) were \textbf{O1} factual conflicts and 234 (56.8\%) were \textbf{O2} framing conflicts. 
\textbf{O1} occurred more often in questions requiring localization of changes or extrema, whereas \textbf{O2} was more common when personas characterized clusters or the strength of a pattern.

\subsection{Case I. From Perceived Patterns to Takeaways}

We compare two public-facing visualizations concerning police violence to examine how agreement among L3 statement pairs relates to agreement among L4 takeaways across personas, as shown in~\autoref{fig:case_study}-I.

For the \#BlackLivesMatter chart, our pipeline selects five personas, spanning political commentary, social-media communication, activism, legal advocacy, and platform analytics.
When asked about the overall trend, all five personas described a decline in supportive sentiment and an increase in neutral sentiment. 
None of the 192 statement pairs was Opposing, although 153 were Disjoint, as the pairs describe different Targets and Patterns.
However, their L4 takeaways differ. For example, the cancel culture critic interprets the decline as evidence of performative conformity, whereas the platform analyst attributes it to a broader mixture of users, aggregators, and automated accounts.
Others interpret the pattern as a sign of shifting public attention, where the organization member treats it as a reason to re-engage neutral audiences, and the lawyer argues for institutional reforms to retain public attention.

The police violence chart presents a complementary relationship between L3 statements and L4 takeaways.
The six selected personas span advocacy, comparative sociology, data analysis, lived experience, law enforcement, and investigative journalism.
When asked about overall distribution over time, the personas emphasize different Target and Pattern of the stacked distribution, including the consistently small Asian/Pacific Islander group and the variable Native American group.
Despite these different emphases in L3 statements, their L4 takeaways converge around a broader concern about racially uneven exposure to police violence.
The advocate and sociologist emphasize systemic discrimination. The Cambodian-American persona expresses that aggregation could obscure vulnerable Southeast Asian communities, while the journalist focuses on the under-recognition of Indigenous deaths. The law enforcement officers also acknowledge racial disparities but emphasize training, de-escalation, and operational reform. 

Through qualitative analysis of these two charts, we show that Compatible statement pairs can support different explanations, while Disjoint statement pairs can contribute to a shared broader concern.
Our TPS framework identifies where the L3 statements align or differ, while the review of L4 takeaways reveals how personas interpret those observations.

\subsection{Case II. Redesigning a Communicative Visualization}
We next examine a public-facing unemployment visualization~\cite{unemployment}, with an alternative design informed by design guidelines~\cite{franconeri2021science}, as shown in~\autoref{fig:case_study}-II.
Both versions show the United States unemployment rate following the 2007--2009 recession, with different visual emphasis: the original states the 8\% unemployment benchmark in the caption, whereas the revision adds a reference line, shaded regions, and repositioned annotations.
We reuse the same questions and the same seven personas across both versions.
The seven selected personas span public administration, political strategy, policymaking, worker experience, macroeconomics, and corporate management.


Our pipeline initially generates two questions about the overall trajectory and a local fluctuation. 
For both questions, the redesign does not yield meaningfully different ambiguity profiles.
We consider that neither question directly evaluates the intended comparison between the observed unemployment rate and the administration's 8\% benchmark. 
This motivates us to add a question that directly assesses the effectiveness of the redesign: \textit{Does this chart make the case that the administration succeeded or failed on its unemployment promise?}

In this newly added question, we observe that the mention of the benchmark increased from 8 of 16 L3 statements in the original design to 13 of 14 in the redesign.
The specific Targets still vary, including the unemployment line, its relationship to the reference line, and the shaded regions, but there are more consistent mentions of the benchmark comparison in the L3 statements in the redesign.
Compatible labels consequently increased from 27 of 109 (24.8\%) to 37 of 84 (44.0\%), while Disjoint labels decreased from 78 of 109 (71.6\%) to 44 of 84 (52.4\%). Opposing labels remained nearly unchanged at 3.7\% and 3.6\%.

The remaining Opposing pairs reveal why the visual emphasis of the benchmark does not imply a uniform interpretation. In the original design, one persona emphasizes the decline from late 2010 through 2012, whereas another describes the unemployment rate hovering near 9\% during 2010–2011.
In the redesign, both sides explicitly reference the visual emphasis of the benchmark, but one persona emphasizes the eventual crossing below 8\%, while another describes the prolonged period above the reference line.
These differences persist in the L4 takeaways. In the redesign, all personas acknowledge that the original deadline is missed, but some interpret the subsequent decline as delayed success. Others mention policy failure, structural labor-market constraints, or worker consequences. 

These results suggest that the redesign changes how chart interpretations are organized rather than eliminating interpretive variation.
In the redesign, more L3 statements anchor the observed unemployment to the benchmark, consequently producing more Compatible labels among statement pairs describing the same Targets. Yet, we also find that personas continue to emphasize different time periods and reach different explanations.
Evaluating a visualization redesign therefore requires examining not only whether its ambiguity profile becomes narrower, but also how chart interpretations change and how those changes are associated with the intended communicative goal of the visualization.

By relating TPS, L3 statements, and L4 takeaways, our methodology systematically distinguishes variation in what personas pay attention to, how they describe it, and what conclusions they draw.
Across the two cases, we observe that ambiguity labels assigned to L3 statement pairs do not alone determine agreement among L4 takeaways.
The redesign case further shows that an ambiguity profile is meaningful only relative to the question and communicative goal being evaluated.
These findings motivate treating the interpretive variation as a visualization design objective and the ambiguity profile as inspectable design feedback, to be elaborated in the following subsection.
%

\section{Discussion}

Our findings position interpretive variation as an evaluation target in its own right. Whereas visualization evaluation often asks whether a chart is accurate, readable, or effective for a task, an ambiguity profile asks how the same visual evidence can support different L3 statements within persona-conditioned L4 takeaways. This shift opens opportunities to make interpretive differences actionable during authoring, to evaluate coverage of possible perspectives, and to incorporate communicative concerns into emerging agentic visualization workflows.

\subsection{Interpretive Range as a Visualization Design Objective}

Our framework suggests that authors can reason about a visualization's \emph{interpretive range}: the set of L4 takeaways that its design makes salient. For an explanatory visualization, the goal may be to concentrate that range around an intended message. For an exploratory visualization, supporting several complementary takeaways may be desirable. The distinction between Disjoint and Opposing is important here: Disjoint statement pairs reveal how attention is distributed across different features, while Opposing statement pairs reveal competing characterizations of the evidence.

The case studies further show that agreement at one interpretive level does not imply agreement at another. Compatible L3 statement pairs can support divergent L4 takeaways, while Disjoint pairs can converge on a shared contextual concern. Increasing agreement about visible patterns may therefore clarify what viewers see without aligning what they conclude. Evaluating interpretive range requires examining both the relationships among L3 statements and the L4 takeaways they support, especially when the intended message concerns explanation, attribution, or action.

\subsection{Ambiguity Profiles Are Conditional on Evaluation Goals}

Our redesign case also shows that an ambiguity profile is not an intrinsic property of a visualization; it is conditional on the interpretive question and communicative goal used to evaluate it. Questions about the overall trend and short-term fluctuation did not reveal whether the redesign clarified the administration's 8\% benchmark. A question aligned with that goal, however, exposed a shift toward Compatible pairs. Evaluation should therefore specify the claim, judgment, or decision that the visualization is intended to support and use multiple questions when different facets of its interpretive range matter.

Communicative visualization research has often evaluated alignment between design intent and audience interpretation by asking intended audiences to articulate the message a visualization conveys, typically through open-ended questions~\cite{knoll2025gulf, schuster2024being}. Recent work has also explored written visualization objectives as explicit representations of design intent that can guide ideation, iteration, and design rationale~\cite{adar2020communicative, stokes2024s}. Our framework connects these two directions by making a stated visualization objective directly comparable with sampled, persona-conditioned L4 takeaways. When the takeaways diverge from the objective, their supporting L3 statements and Target–Pattern–Scope descriptions help identify which features, patterns, or scopes produced that divergence. 

This view turns ambiguity from a binary defect into a design variable that can be examined against communicative intent. An authoring tool could accept an intended takeaway, compare it with sampled L4 takeaways and their supporting L3 statements, and show whether revisions narrow, broaden, or redirect the ambiguity profile. Our ambiguity profiles could thereby complement audience-based evaluation by providing structured feedback during design iteration, before or alongside studies with intended audiences. The taxonomy provides the vocabulary needed to make those changes visible without assuming that every alternative reading should be eliminated.

\subsection{Ambiguity Profiles as Inspectable Design Feedback}

An ambiguity profile offers an intermediate representation between a single quality score and a collection of unstructured model responses. The \hlN{Compatible}, \hlD{Disjoint}, and \hlO{Opposing} ambiguity labels summarize what kind of variation occurs, while Target--Pattern--Scope identifies where it occurs. This structure enables an author to move from the overall distribution to representative statement pairs, their specific differences, and the L4 takeaways they support. Lexara similarly demonstrates the value of interpretable metrics and multi-level exploration for inspecting automated evaluations \cite{palani2026lexara}.

The ambiguity labels also suggest different design actions. Disjoint Targets or Scopes can prompt authors to reconsider titles, annotations, or visual emphasis, whereas Opposing Patterns can prompt inspection of the selected interval, baseline, scale, or framing. Combined with VLM-based design feedback \cite{shin2025visualizationary}, these explanations could support cycles of revision and re-evaluation. A visualization badge could surface potential ambiguity alongside a published chart \cite{edelsbrunner2025badges}, with the ambiguity profile and example L4 takeaways providing the evidence behind the badge.

\subsection{From Synthetic Users to Perspective Sampling}

Persona-conditioned simulation is often framed as a way to approximate how users might respond. Our method instead uses personas to expand the interpretation space that an evaluation explores. Topic-based retrieval makes the probes relevant to a chart, diversity sampling reduces repetition, and pairwise analysis of L3 statements preserves differences that would disappear if the L4 takeaways were averaged into a single synthetic audience judgment.

This reframing introduces \emph{perspective coverage} as a useful evaluation criterion: whether a sampling procedure surfaces materially distinct ways of reading the chart. The Target--Pattern--Scope framework makes that coverage inspectable rather than equating diversity with surface-level differences in wording. The relevant output is the range and structure of sampled L4 takeaways and L3 statements, not an estimate of their prevalence among real viewers. This use of personas remains sensitive to which attributes are elicited \cite{hu2024persona}; questions of representation and correspondence with human audiences remain important directions for validation \cite{agnew2024inclusion}.

\subsection{Ambiguity Evaluation in Agentic Workflows}

Recent agendas for agentic visualization envision systems that generate visualizations, test them through simulated human feedback, and iteratively improve their designs \cite{zhu2026agentsfuturevis}. Our work contributes a concrete evaluation representation for such workflows. Rather than returning a generic critique, an ambiguity evaluator produces L4 takeaways and supporting L3 statements, assigns ambiguity labels to statement pairs, and explains those labels through Target--Pattern--Scope.

These outputs could close a feedback loop in which an agent generates design alternatives, evaluates their ambiguity profiles, proposes targeted revisions, and shows the author how each revision changes the ambiguity profile. Such a workflow gives the agent a role in searching both the design space and the interpretation space, while the author specifies the intended message and decides which differences matter. Our framework thus offers a concrete pattern for mixed-initiative evaluation centered on communicative consequences, extending agentic visualization beyond generation and technical correctness.

\subsection{Limitations}



Our pipeline's dependence on VLMs has inherent limitations. Selecting the highest-performing models minimizes, but does not eliminate, chart-reading errors. Reasoning errors in statement-pair labeling also occur due to the novelty of the task. The applicability of our pipeline and taxonomy to chart types beyond those evaluated is also limited, as visualization-literacy benchmarks and chart datasets are heavily weighted toward bar and line charts. VLM output also depends on prompting, stochastic generation, and model behavior, which may not reflect human reasoning.


The use of PersonaHub \cite{personahub} and our clustering process only provides semantic coverage. Since persona selection involves minimal reasoning, it is possible to exclude personas that are not semantically related to the topic but would be interested in the topic in the real world, and vice versa. This can lead to biased ambiguity profiles.


We acknowledge that the 115 human-labeled statement pairs constitute a limited sample and that confidence in each reference label may vary due to the flexibility of natural language. The few-shot examples provided in the labeling prompt are also based on this limited set.


In this work, we limited our scope to static, isolated visualizations, and thus do not account for ambiguities introduced via animation or interaction, or publication context. Visualizations requiring external knowledge for interpretation were also out of scope for this work. 


With our design, the number of persona-conditioned chart-interpretation calls scales linearly with the number of interpretive questions and personas. However, the raw number of statement pairs scales quadratically with both the number of personas and the average number of L3 statements per persona, which can make large-scale testing costly. Our experiments suggest that performance-tier models are more likely to generate more L3 statements per persona than faster models.


While we present our framework as an additional tool for visualization authors, we caution against fully replacing synthetic feedback with human feedback.

\section{Conclusion}

In this work, we define and address interpretive ambiguity, the phenomenon in which individuals derive varying takeaways from the same visualization. We introduce a model-agnostic pipeline that generates persona-conditioned L4 takeaways and supporting L3 statements to construct an ambiguity profile. This profile reports the distribution of Compatible, Disjoint, and Opposing ambiguity labels assigned to statement pairs using our Target--Pattern--Scope framework and ambiguity taxonomy. Through case studies involving charts of controversial topics, we show how the pipeline can identify not only opposing views from synthetic personas with competing interests but also areas that visualization authors and distributors should handle with more care.
More broadly, our work positions AI not only as a visualization generator but also as a tool for anticipating how designs may be interpreted. As AI enters data-communication workflows, such evaluation can inform iteration while authors retain responsibility for consequential judgments and audience engagement.


\bibliographystyle{abbrv-doi-hyperref}
\vspace{0.02in}
\bibliography{reference}

\clearpage
\appendix
\section{Ambiguity-Profile Pipeline Implementation Details}
\label{supp:pipeline-details}

This supplement records the lower-level implementation choices and parameter settings for the four-stage ambiguity-profiling pipeline described in Sec. 3.3. Chart profiling and chart interpretation use the chart image, whereas persona selection and statement-pair labeling operate on text. All model calls are stateless and independent.

\subsection{Analytical Boundaries}

The pipeline adopts four boundaries when constructing and interpreting ambiguity profiles. First, its analysis is restricted to L3 statements intended to be supportable from visible chart evidence without requiring external knowledge. Second, it compares statements only when they concern the same chart and respond to the same interpretive question. Third, a pair label characterizes how two statements relate semantically; it does not independently establish that either statement is factually correct, so chart comprehension and statement grounding remain prerequisites. Finally, an ambiguity profile is a diagnostic representation of potential interpretive variation, not an estimate of interpretation prevalence or a definitive binary judgment that a chart is ambiguous.

\subsection{PersonaHub Pool Construction}

We begin with the 200,000 persona descriptions in PersonaHub \cite{personahub}. We apply an ASCII-only heuristic to remove 1,462 descriptions likely to be written in languages other than English. Using LLM-as-a-judge, we also exclude 3,810 descriptions of animals, organizations, or otherwise non-human entities. The resulting pool contains 194,728 personas.

\subsection{Chart Profiling}

We use GPT 5.6 Terra as a fixed upstream model to produce one chart profile shared by all subsequent stages. Because model routing depends on the chart type recorded in the profile's L1 statement, chart-type-specific routing begins only at chart interpretation.

The model records the chart's title or caption, chart type, axes, variables, encodings, and categories in an L1 statement. Because this statement is constant across personas and questions, we extract it once rather than asking each persona-conditioned model call to repeat it. The model also generates a single phrase describing the chart's real-world topic. Rather than copying an isolated title, caption, or axis label, it considers the complete chart so the topic can support retrieval of personas with relevant interests or backgrounds.

The model considers six analysis tasks associated with L3 statements in existing taxonomies \cite{lee2006task,stokes2025writerankratecomparing}:

\begin{itemize}
    \item \textbf{Characterize Distribution:} Describe the overall shape or spread of a quantitative value across data cases.
    \item \textbf{Cluster:} Identify groups of data cases with similar attribute values.
    \item \textbf{Correlation:} Examine whether two attributes appear to have a meaningful relationship.
    \item \textbf{Large Trend:} Describe the overall direction across the full dataset.
    \item \textbf{Small Trend:} Describe short-term changes within the dataset.
    \item \textbf{Find Anomalies:} Identify outlier trends in the chart.
\end{itemize}

It selects only the tasks supported by the chart's visible structure, provides brief visual evidence for each selection, and generates one open-ended, chart-specific question per selected task. This chart-level selection avoids imposing a fixed chart-type--task mapping on structurally different charts that share a broad type. Unlike L1 or L2 lookup questions, the generated questions permit variation in which visible patterns viewers foreground and how they characterize them. Unlike L4 questions, they do not require external knowledge and can therefore be checked against the chart. Asking every persona the same generated question controls the comparison because pairwise relationships are calculated only among L3 statements responding to that question.

\subsection{Persona Retrieval and Threshold Selection}

Let $\mathcal{P}$ be the persona pool after filtering PersonaHub. We convert every description into a 1,536-dimensional embedding using OpenAI's \texttt{text-embedding-3-small}; each $\mathbf{p}\in\mathcal{P}$ contains the persona's textual description and its raw embedding. The chart topic is embedded in the same space, and personas are ranked by cosine similarity to the embedded topic.

We determine a chart-specific similarity threshold $\Theta_\mathrm{final}$ with an LLM-as-a-judge and the binary search shown in \autoref{alg:persona-threshold}. Personas at or above the threshold form $\mathcal{P}_{\mathrm{selected}}$. For each chart, we L2-normalize the candidate embeddings, fit an uncentered singular value decomposition (SVD) to reduce them to 50 dimensions, and L2-normalize the reduced embeddings again. Pilot work with 30 and 100 dimensions yielded similar results, so we retained 50 as the default.

\begin{algorithm}[t]
\caption{Persona Threshold Selection}
\label{alg:persona-threshold}
\begin{algorithmic}[1]
\Require Topic $T$, ranked personas with similarity scores, tolerance $\epsilon=0.01$
\State $L \gets 0.0$
\State $U \gets 1.0$
\State $\Theta_{\mathrm{final}} \gets \mathrm{unset}$
\State $t \gets 0$
\While{$U - L > \epsilon$ and $t < 20$}
    \State $P \gets$ untested persona with similarity score closest to $(L+U)/2$
    \If{no untested persona remains}
        \State \textbf{break}
    \EndIf
    \State $\textit{relevant} \gets \textsc{LLMJudge}(P, T)$
    \If{$\textit{relevant}$}
        \State $U \gets \mathrm{score}(P)$ \quad \textit{// Include more personas}
        \State $\Theta_{\mathrm{final}} \gets \mathrm{score}(P)$
    \Else
        \State $L \gets \mathrm{score}(P)$ \quad \textit{// Include fewer personas}
    \EndIf
    \State $t \gets t+1$
\EndWhile
\If{$\Theta_{\mathrm{final}}=\mathrm{unset}$}
    \State $\Theta_{\mathrm{final}} \gets U$
\EndIf
\State $\mathcal{P}_{\mathrm{selected}} \gets \{P_i \mid \mathrm{score}(P_i) \geq \Theta_{\mathrm{final}}\}$
\State \Return $\Theta_{\mathrm{final}}, \mathcal{P}_{\mathrm{selected}}$
\end{algorithmic}
\end{algorithm}

The search is calibrated independently for each chart over the cosine-similarity interval $[0,1]$, with tolerance $0.01$ and at most 20 judgments. GPT 5.6 Terra judges whether each tested persona is relevant to the chart topic. The search assumes that personas become monotonically more relevant as similarity increases; the implementation does not separately test this assumption. If judgments fluctuate, it retains the most recently observed relevant score. If no tested persona is judged relevant, it uses the current upper bound.

\subsection{Persona Clustering}

We use two complementary estimates to determine how many groups to form among the reduced candidate embeddings. First, we calculate pairwise cosine distance,

\begin{equation}
    d(\mathbf{p}_i,\mathbf{p}_j)=1-\cos(\mathbf{p}_i,\mathbf{p}_j),
\end{equation}

and construct an average-linkage dendrogram. Let $\tau_\mathrm{den}$ be the median pairwise cosine distance. We cut the dendrogram at $\tau_\mathrm{den}$ and define $k_\mathrm{natural}$ as the number of remaining groups. When more than 5,000 personas match a chart, we estimate $k_\mathrm{natural}$ from a random sample of 5,000 to avoid the quadratic memory cost of the distance matrix; all matched personas remain in the subsequent $k$-means clustering.

Second, we calculate $k_\mathrm{max}$, the largest number of $k$-means clusters that remains reproducible under perturbations of $\mathcal{P}_{\mathrm{selected}}$. For each $k_\mathrm{candid}\in K=\{2,3,4,5,6,8,10,12,16,20\}$, we:

\begin{enumerate}
    \item Fit $k_\mathrm{candid}$-means to $\mathcal{P}_{\mathrm{selected}}$ as the reference partition.
    \item Draw 20 subsamples, each containing 80\% of $\mathcal{P}_{\mathrm{selected}}$, without replacement.
    \item Fit $k_\mathrm{candid}$-means independently to each subsample.
    \item Compare each subsample partition with the corresponding reference assignments using the Adjusted Rand Index (ARI) \cite{hubert1985ari}.
    \item Average the 20 ARI values:
    \begin{equation}
        A(k)=\frac{1}{20}\sum_{b=1}^{20}
        \mathrm{ARI}\!\left(z^{\mathrm{full}}_{I_b},z^{(b)}_{I_b}\right).
        \label{eq:supp-ari}
    \end{equation}
\end{enumerate}

We define $k_{\max}=\max\{k\in K:A(k)\geq0.6\}$. The candidate grid, 20 subsamples, 80\% subsample size, and 0.6 ARI threshold were selected through a pilot investigation of 10 charts rather than derived from first principles. The same pilot supported uncentered SVD and $k$-means: uncentered SVD preserved the cohesion of the original embeddings, and $k$-means was more reproducible under subsampling than the tested alternatives. If no $k$ meets the 0.6 threshold, $k_\mathrm{max}=0$.

We set $k_\mathrm{selected}=\min(k_\mathrm{natural},k_\mathrm{max})$ and fit $k$-means++ with 10 initializations, at most 300 iterations, and random seed 42. From each cluster, we return its medoid: the actual persona with the greatest mean cosine similarity to the other cluster members.

We flag a chart as degenerate when fewer than 20 personas match or when $k_\mathrm{max}\leq2$; in these cases, the automatic pipeline stops rather than guessing a cluster count. Across the nine nondegenerate charts in our pilot, $k_\mathrm{selected}$ yielded three to six representative personas. The candidate grid caps automatic selection at 20 representatives, but we impose no cap on the number of matched personas assigned during $k$-means.

\subsection{Chart Interpretation}

For a given chart, chart interpretation makes $|Q|\times|P|$ VLM calls, where $Q=\{q_1,\ldots,q_U\}$ is the set of generated questions and $P=\{p_1,\ldots,p_N\}$ is the set of representative personas. Each call receives the chart image, one question, the chart's L1 statement, the unmodified PersonaHub description, and a prompt adapted from Charts-of-Thought \cite{das2025charts}. The VLM first answers from the assigned persona's perspective by producing an L4 takeaway, then identifies L2 observations and chart-grounded L3 statements that support it. In pilot experiments, eliciting the L4 takeaway before its L3 statements produced statements that were more responsive to the question and less likely to incorporate information not visible in the chart.

We do not prescribe the number of L3 statements. In our runs, the VLM returned between 1 and 7 per call (median 3, SD 1.16). In the same image-grounded call, the model identifies the Target, Pattern, and Scope of every L3 statement for use in statement-pair labeling. The prompt includes extraction, sorting, and verification steps, but we do not run a separate model to verify each L3 statement against the chart.

\subsection{Statement-Pair Construction and Scaling}

The final stage constructs within-question, between-persona statement pairs. For selected personas $P=\{p_1,p_2,\ldots,p_N\}$, let $S_{q,n}=\{s_{q,n}^{(1)},s_{q,n}^{(2)},\ldots,s_{q,n}^{(R_{q,n})}\}$ be the L3 statements generated for persona $p_n$ and question $q\in Q$. The number of pairs for a question is

\begin{equation}
    L=\sum_{i=1}^{N}\sum_{j=i+1}^{N}|S_{q,i}||S_{q,j}|.
    \label{eq:supp-pair-count}
\end{equation}

Restricting comparisons to one question reduces irrelevant pairings because different generated questions may ask about different variables or patterns. Each pair is passed to the relationship-labeling protocol described in Sec. 5.4. The number of chart-interpretation calls scales quadratically with both the number of representative personas and the average number of L3 statements per persona.

\subsection{Additional Ambiguity-Taxonomy Examples}

The main paper uses representative examples to explain the boundaries among the three top-level labels. The following pairs document additional subtypes from Sec. 5.3.

\textbf{Equivalent (EQ).} The following statements use different wording but have equivalent Target, Pattern, and Scope in the context of a chart whose displayed age range is 15 to 85:

\vspace{0.5em}
\noindent \textbf{A:} \textit{\tpsScope{As individuals age from 15 to 85}, \tpsTarget{the amount of time they spend alone} \tpsPattern{generally increases}.}
\vspace{0.3em}

\noindent \textbf{B:} \textit{\tpsTarget{Time spent alone} \tpsPattern{generally increases with age} \tpsScope{across the entire lifespan shown}.}
\vspace{0.5em}

\textbf{Directional entailment.} In the time-use chart in Fig. 2-I, the statements share their Target and Pattern, while \textbf{B}'s Scope is contained within \textbf{A}'s:

\vspace{0.5em}
\noindent \textbf{A:} \textit{\tpsTarget{Time spent alone} \tpsPattern{increases steadily as age increases} \tpsScope{from 15 to 85}.}
\vspace{0.3em}

\noindent \textbf{B:} \textit{\tpsTarget{Time spent alone} \tpsPattern{increases steadily as age increases} \tpsScope{from 20 to 85}.}
\vspace{0.5em}

An increase from age 15 to 85 guarantees an increase from age 20 to 85, but not conversely. The pair is therefore Compatible through directional entailment from \textbf{A} to \textbf{B}.

\textbf{Complementary parts (D1).} These statements concern different parts of the same response distribution, but neither entails or conflicts with the other:

\vspace{0.5em}
\noindent \textbf{A:} \textit{The opinion is \tpsPattern{primarily split between} \tpsTarget{those who believe the U.S. has no responsibility and those who believe it does}.}
\vspace{0.3em}

\noindent \textbf{B:} \textit{\tpsPattern{A small portion of} \tpsTarget{the public is undecided on the U.S. responsibility in Iraq}.}
\vspace{0.5em}

\textbf{Mixed TPS (D5).} In this pair, \textbf{A}'s Target is broader than \textbf{B}'s, while \textbf{B}'s Pattern is more specific than \textbf{A}'s. Because the component-level specificity relations point in different directions, neither complete statement entails the other:

\vspace{0.5em}
\noindent \textbf{A:} \textit{\tpsTarget{The percentage of critically ill COVID-19 patients for both sexes} \tpsPattern{decreases} \tpsScope{in the 70--79 and 80+ age groups}.}
\vspace{0.3em}

\noindent \textbf{B:} \textit{\tpsTarget{The percentage of critically ill females with COVID-19} \tpsPattern{drops sharply} \tpsScope{after the 60--69 age group}.}
\vspace{0.5em}

\section{Dataset Curation and VLM Benchmark Details}
\label{supp:dataset-model-details}

\subsection{Dataset Selection and Curation}

We reviewed visualization-literacy and chart-question-answering datasets for chart-type coverage and design variety. We sought real-world, public-facing visualizations with conventional chart structures and minimal embellishment, paired with questions that directly test chart comprehension. We excluded datasets described as containing synthetic data. We did not screen every chart for compliance with established design guidelines \cite{lan2024came,lo2022misinformed}; the resulting corpus should therefore not be interpreted as a verified collection of well-designed visualizations. We considered the chart types represented in the VLAT \cite{vlat}: line, bar, stacked bar, 100\% stacked bar, pie, histogram, scatterplot, bubble, area, stacked area, choropleth, and treemap. Infographics, dashboards, and other multi-chart visualizations were outside our scope.

ChartMuseum \cite{tang2026chartmuseum} and ChartQAPro \cite{masry2025chartqapro} best met these criteria. We selected ChartQAPro after closer inspection because a larger share of its charts fell within the VLAT types. Its 1,948 chart--question pairs contained 1,252 unique chart images after we removed 696 duplicates by exact comparison of their base64 encodings.

ChartQAPro does not provide chart-type metadata, so a VLM council comprising GPT 5.4 Nano, Gemini 3.5 Flash-Lite, and Claude Haiku 4.5 independently assigned each unique chart one of the VLAT types or one of three additional categories. We used \emph{bar + line} for overlaid bar and line marks, \emph{infographic/dashboard} for multi-chart designs or charts with extensive annotations or captions, and ``other'' as a fallback. Majority vote categorized 1,200 of the 1,252 charts; members of the research team manually categorized the remaining 52. To avoid sparse categories with closely related structures, we merged histograms with bar charts, 100\% stacked bar charts with stacked bar charts, and bubble charts with scatterplots.

Because ChartQAPro contained relatively few pie charts and scatterplots, we supplemented it with 24 pie charts from ChartQA \cite{masry2022chartqa} and 5 pie charts and 24 scatterplots from ChartMuseum. After removing 492 infographics/dashboards, 24 ``other'' charts, 6 choropleths, and the empty treemap category, 730 ChartQAPro charts remained. The 53 supplemental charts produced a final corpus of 783 unique charts.

For the chart--question benchmark, we retained ChartQAPro's ``factoid'' free-response, ``multi-choice'' A--D, and ``fact checking'' True/False questions for charts in the final corpus. We excluded ``conversational'' entries that package multiple questions for one chart, ``hypothetical'' questions requiring predictions, and questions marked ``unanswerable.'' This procedure produced 689 chart--question pairs across eight chart types.

\subsection{Candidate Model Coverage}

We purposively selected a non-exhaustive set of nine recent proprietary VLMs: one model from each of the GPT\footnote{https://developers.openai.com/api/docs/models}, Gemini\footnote{https://ai.google.dev/gemini-api/docs/models}, and Claude\footnote{https://platform.claude.com/docs/en/about-claude/models/overview} families at each of three provider-described tiers---fast, balanced, and performance. These tier labels organize the comparison and are not treated as analytical categories. We ran the benchmark in August 2026. Reported costs use recorded API token counts and provider per-token prices at the time of evaluation.

\subsection{Benchmark Procedure}

For every retained chart--question pair, each candidate model received a Charts-of-Thought-based prompt \cite{das2025charts}, the question, and the chart image. The prompt instructed the model to extract chart values and information, sort the extracted values, verify them against the chart, and answer the question. We enforced structured JSON output and a 10,000-token output budget, inclusive of reasoning tokens where reported by the provider, and otherwise used provider defaults. Failed calls were retried up to four times; after five unsuccessful attempts, we recorded an empty answer.

Gemini 3.5 Flash-Lite served as an LLM judge. Semantic judging accommodated formatting differences such as percentages represented as 4, 0.04, 4\%, or within a sentence. The judge received the ChartQAPro reference answer and anonymized model responses. We selected this judge because it was a recent, low-cost model suitable for evaluation at this scale. Temperature was 0, output was capped at 500 tokens, answers involving years received a one-year tolerance, and other numerical answers received a 5\% tolerance. Unintelligible responses, malformed judge outputs, and model calls that failed all five attempts were counted as incorrect. We computed overall accuracy over all 689 retained pairs rather than averaging chart-type accuracies.

When multiple models tied for the highest accuracy on a chart type, we selected the model with the higher overall benchmark accuracy. For a chart type absent from the benchmark, the pipeline defaults to Gemini 3.6 Flash, the model with the highest overall accuracy. Routing applies only during chart interpretation; chart profiling uses the fixed model described in Sec. 3.3.1.

\end{document}